\documentclass{elsart}
\journal{Optics Communications}

\usepackage{graphicx,color}
\usepackage{epsfig}

\usepackage{dcolumn}
\usepackage{bm}
\usepackage{amssymb}
\usepackage{array}

    \newcommand{\ket}[1]{\ensuremath{|\,{#1}\,\rangle}}
    \newcommand{\bra}[1]{\ensuremath{\langle\,{#1}\,|}}
    \newcommand{\braket}[2]{\ensuremath{\langle\,{#1}\,|\,{#2}\,\rangle}}

    \newcommand{\lsub}[1]{\ensuremath{_{_{\!\scriptstyle #1}}}}

    \renewcommand{\vec}[1]{\mbox{\boldmath{\ensuremath{{#1}}}}}

    \newcommand{\itgf}[1]{\ensuremath{\int\!\!d{#1}\,}}

    \newcommand{\sinc}{\ensuremath{\mbox{\hspace{1.3pt}sinc}\,}}

\begin{document}

\begin{frontmatter}
\title{Generating mixtures of spatial qubits}

\author[UC,UFMG]{G. Lima},
\ead{glima@udec.cl}
\author[UC]{F. A. Torres-Ruiz},
\author[UFMG]{Leonardo Neves},
\author[UC]{A. Delgado},
\author[UC]{C. Saavedra},
\author[UFMG]{S. P\'adua}
\address[UC]{Center for Quantum Optics and Quantum Information,
Departamento de F\'{\i}sica, Universidad de Concepci\'on,  Casilla
160-C, Concepci\'on, Chile.}
\address[UFMG]{Departamento de F\'{\i}sica, Universidade Federal de
Minas Gerais, Caixa Postal 702, Belo~Horizonte,~MG 30123-970,
Brazil.}

\begin{abstract}
In a recent letter [Phys. Rev. Lett. 94, 100501 (2005)], we
presented a scheme for generating pure entangled states of spatial
qudits ($D$-dimensional quantum systems) by using the momentum transverse
correlation of the parametric down-converted photons. In this work we discuss a
generalization of this process to enable the creation of mixed states. With the technique
proposed we experimentally generated a mixture of two spatial qubits.
\end{abstract}

\begin{keyword}
spatial qubits states, spatial qudits states, mixed states
\PACS 42.50.Dv, 03.67.Bg
\end{keyword}

\end{frontmatter}

\section{Introduction}
\label{sec:intro}

The experimental generation of pure entangled states and their entanglement properties
have been extensively explored \cite{Kwiat1,Kwiat2,Kwiat3,zeilinger2,gisin2,Leonardo,GLima,howell}.
However, the generation and the quantum properties of mixed states have received much less attention,
even though this is a very important problem. For instance, the recent experimental works done on the
generation of mixtures of two polarized qubits (2-dimensional quantum systems) \cite{White,Kwiat4,Martini}
showed, surprisingly, that for the same state purity there is a class of states which are more
entangled than the Werner states \cite{Werner}.

Besides being a valuable tool that can bring highlights to our understanding of quantum mechanics,
the controlled generation of mixed states can be seen as a technique for state engineering. As was
demonstrated in Ref \cite{White}, partially entangled states can be obtained by introducing a
controlled decoherence into pure maximally entangled states. Another perspective is that,
in some experimental situations, like in quantum key distribution with polarized entangled photons,
the loss of coherence of the quantum state is an intrinsic phenomenon and a limiting constraint
\cite{zeilinger1}. Therefore, it would be of paramount importance if the studies of mixing states
could show a solution to minimize the decoherence in these experiments.

In a recent work \cite{Leonardo,GLima} we showed, theoretically and experimentally, that one can
use the photon pairs created by spontaneous parametric down-conversion to generate pure entangled
states of $D$-dimensional quantum systems. It was the first demonstration of high-dimensional
entanglement based on the intrinsic transverse momentum entanglement of type-II down-converted photons.
Because the quantum space of these photons is defined by the number of different available ways
for their transmission through apertures placed in their paths, we call them spatial qudits.
In this present work, we discuss how a modification of the setup employed to create pure qudit
entangled states can be used to generate mixed states. By using the proposed technique we generated
mixed states of spatial qubits. Up to our knowledge this is the first time that non-polarized mixed
states of qubits are generated. Besides this, we believe that this is also the first time that
a simple experimental technique that allows the generation of mixed states of qudits is discussed.

\section{Theory}
\label{sec:theory}

\subsection{The state of the down-converted photons transmitted through multi-slits}

\begin{figure}
\centerline{\includegraphics[width=0.2\textwidth]{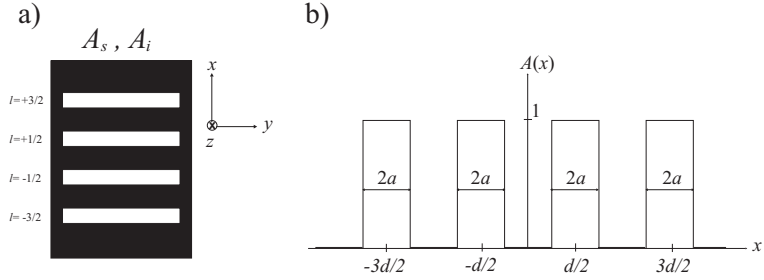}}
\caption{(a) Sketch of a 4-slits aperture. On the left side are shown the slit labels ($l$).
(b) Transmission function of this aperture.
\label{fendas}}
\end{figure}

Spontaneous parametric down-conversion is a nonlinear
optical process where one photon from the pump (p) laser beam
incident to a crystal can originate two other photons, signal (s)
and idler (i)\cite{MandelBook}. The generated photon pairs are
also called twin photons for being generated simultaneously with a
very small temporal uncertainty \cite{HOM-interferometer}. In references \cite{Leonardo,GLima}
we showed that the state of the parametric
down-converted photons, when transmitted through identical multi-slits, with $d$ being
the distance between two consecutive slits, and $a$ the half width of the slits
(See Fig \ref{fendas}), can be written as

\begin{equation} \label{stgeral}
\ket{\Psi} = \sum_{l=-l_D}^{l_D} \; \sum_{m=-l_D}^{l_D}
W_{lm} \; e^{i\frac{kd^{2}}{8z_{A}}(m-l)^2} \; \ket{l}\lsub{s}
\otimes \ket{m}\lsub{i},
\end{equation} where $k$ is the wave number of the pump beam used to generate the twin
photons and $l_D = (D-1)/2$. $D$ is the number of slits in these apertures.
The function $W_{lm}$ is the spatial distribution of the pump beam at the plane of
the multi slits ($z=z_{A}$) and at the transverse position $x=(l+m)d/2$

\begin{equation} \label{wlm}
W_{lm} = W \left[\frac{(l+m)d}{2};z_A\right].
\end{equation}

The $\ket{l}$ (or $\ket{m}$) state is a single-photon state
defined, up to a global phase factor, by

\begin{equation}      \label{base}
\ket{l}\lsub{j} \equiv \sqrt{\frac{a}{\pi}} \itgf{q_{j}}
e^{-iq_{j}ld}\sinc(q_{j}a)\ket{1q_{j}},
\end{equation} and represents the photon in mode $j$ ($j = i, s$)transmitted by
the slit $l$.  $\ket{1\vec{q}_{j}}$ is the Fock state for one
photon in mode $j$ with transverse wave vector $\vec{q}_{j}$. The base $\{\,\ket{l}\lsub{j}\}$
satisfies $\lsub{j}\braket{l}{l'}\lsub{j}=\delta_{ll'}$. We use
these states to define the logical states of the qudits and thus,
it is clear that Eq.~(\ref{stgeral}) represents a composite system
of two qudits. Each qudit is represented by a vector in a Hilbert space of
dimension $D$ because the degrees of freedom of each photon are the $D$
available paths for their transmission through the multi-slits.

It can be seen from Eq.~(\ref{stgeral}) and Eq.~(\ref{wlm}) that it
is possible to create different pure states of spatial qudits if one
knows how to manipulate the pump beam in order to generate distinct
transverse profiles at the plane of the multi-slits. In Ref. \cite{GLima},
for example , we showed experimentally that an entangled state of spatial ququads ($D = 4$)

\begin{equation}        \label{qudits}
\ket{\Psi} = \frac{1}{2} \sum_{l=-\frac{3}{2}}^{\frac{3}{2}}
e^{ik\frac{d^{2}l^{2}}{2z_{A}}} \; \ket{l}\lsub{s} \otimes
\ket{-l}\lsub{i},
\end{equation} can be generated when the pump beam is focused at the plane of two
identical four-slits in such a way that it is non-vanishing except
in a region smaller than the dark part of these apertures.

\subsection{Controlled generation of mixed states}
\label{sec:generation}

Suppose now that, before reaching the crystal, the pump beam
passes through an unbalanced Mach-Zehnder interferometer where the
transverse profile of the laser beam is modified differently in
each arm. If the difference between the lengths of these arms is
set larger than the laser coherence length, we will obtain an
incoherent superposition of the twin photon states generated by each arm.
Whenever multi-slits are placed in the propagation paths of these photons,
we will be generating a mixed state of two spatial qudits.

\begin{figure}[tbh]
\begin{center}
\rotatebox{-90}{\includegraphics[width=0.5\textwidth]{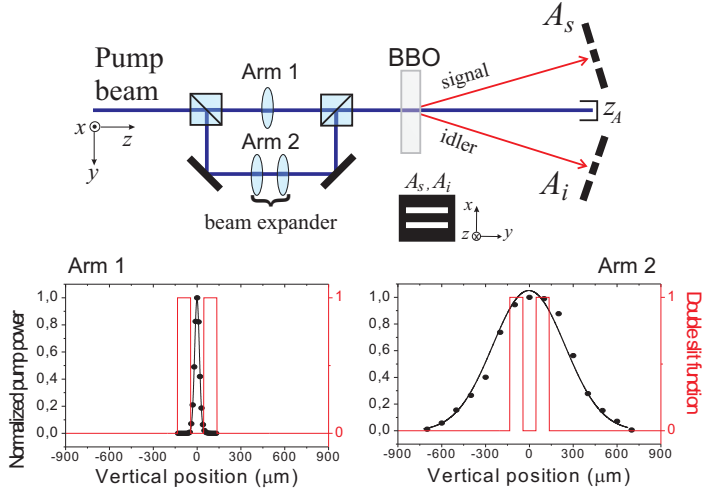}}
\end{center}
\caption{Schematic diagram of the experimental setup used for generating mixed states of spatial qubits.
The pump beam that crosses arm 1 is focused at the plane of the double slits ($z=z_A$). When it passes
through arm 2, it will be in a broader spatial region at this same plane. The transverse profiles
generated are illustrated and compared with the double slit's transmission function. $A_s$ and $A_i$ are the
double-slits in signal and idler propagation paths, respectively.} \label{setup}
\end{figure}

To generate a mixed state of two spatial qubits we performed
the experiment outlined in Fig \ref{setup}. A $5$~mm
$\beta$-barium borate crystal is pumped by a $500$~mW krypton
laser emitting at $\lambda = 413$~nm for generating SPDC. Before
being incident upon the crystal, the pump beam crosses an unbalanced
Mach-Zehnder interferometer. The difference between the lengths of
the interferometer arms ($200$~mm) is set larger than the laser
coherence length ($80$~mm). Two identical double slits $A_s$ and
$A_i$ are aligned in the direction of the signal and idler beams,
respectively, at a distance of $200$~mm from the crystal ($z_A$).
The slits' width is $2a = 0.09$~mm and their separation, $2d =
0.18$~mm. At arm 1 of the interferometer we place a lens that
focuses the laser beam at the plane of these double slits into a
region smaller than $d$. In arm 2, we use a set of lenses that
increases the transverse width of the laser beam at $z_A$. The
transverse profiles generated are illustrated in
Fig ~\ref{setup}. The photons transmitted through the
double-slits are detected in coincidence between the detectors
$D_i$ and $D_s$. Two identical single
slits of dimension $5.0$ x $0.1$ mm and two interference filters
with $8$ nm full width at half maximum (FWHM) bandwidth are placed
in front of the detectors.

Using Eq.~(\ref{stgeral}) and Eq.~(\ref{wlm}), we can show that
the two-photon state, after the double slits, when only arm 1 is
open, is given by
\begin{equation}        \label{qubittomo}
\ket{\Psi}_1 =
\frac{1}{\sqrt{2}}(\ket{+}_s\ket{-}_i+\ket{-}_s\ket{+}_i).
\end{equation}

To simplify, we used the state $\ket{+}_j$ ($\ket{-}_j $) to
represent the $j$ photon being transmitted by the upper (lower)
slit of the respective double slit (i.e., $l=+,-$). The state shown in
Eq.~(\ref{qubittomo}) is a maximally entangled state of two
spatial qubits.

However, if the laser beam crosses only arm 2, the
state of the twin photons transmitted by the apertures will be
\begin{eqnarray}        \label{prodtomo}
\ket{\Psi}_2 &= &\frac{1}{2}
\,e^{i\phi}(\ket{-}_s\ket{+}_i+\ket{+}_s\ket{-}_i)
\nonumber \\
& &+ \frac{1}{2}(\ket{-}_s\ket{-}_i + \ket{+}_s\ket{+}_i),
\end{eqnarray} where $\phi = \frac{k d^2}{8 z_{A}}$. The state $\ket{\Psi}_2$ is
just partially entangled and, as was calculated in Ref \cite{ConcEsp}, it has a concurrence \cite{Wootters}
which is about three times weaker than the concurrence of the state $\ket{\Psi}_1$.

Therefore, the two-photon state generated in our experiment, when
the two arms are liberated, is a mixed state of the spatial
maximally entangled state of Eq.~(\ref{qubittomo}) and the state
of Eq.~(\ref{prodtomo}). It is described by the density operator

\begin{equation}        \label{OPDENSTOMO}
\rho = A \ket{\Psi}\lsub{\,1\,1\!}\bra{\Psi} + B
\ket{\Psi}\lsub{\,2\,2\!}\bra{\Psi},
\end{equation} where $A$ and $B$ are the probabilities for generating
the states of arm 1 and arm 2, respectively.

It is interesting to note that this generation can be completely controlled. Besides the fact that we can
control the states generated in each arm (by controlling for each arm the pump beam transverse profile generated at the plane of the multi-slits), we also can control the probabilities $A$ and $B$ for generating the twin photon states. This is done by controlling the amount of pump power sent through each arm of the interferometer. For this purpose, we can replace the beam splitter at the entrance of the
Mach-Zehnder interferometer by a polarizer beam splitter. Hence, a half wave
plate in front of the pump beam before the interferometer allows
us to propagate defined polarizations along the arms in such way that a rotation
of the HWP behaves as a mechanism for a fine control of the fraction of the pump power
sent through each arm. Furthermore, a HWP rotated at $45^{\circ}$ is inserted in
the arm with horizontal polarization, so that photons arriving at the crystal
have vertical polarization. This mechanism allows for pumping the non-linear crystal
with a constant pump power.

\section{Results and Discussion}
\label{sec:discussion}

The theory developed in the previous two sections, besides being quite appealing, is
straightforward and now we show that our experimental results are in strong agreement with it.
We stress that the subject of state determination \cite{Vogel1,Walmsley,Vogel2,Smithey,Vogel3,James}
of photonic states that are not defined in terms of the photon's polarization is not trivial and,
for this reason, the results here presented should be seen as a step forward in this study for spatial mixed
states.

\subsection{The probability of the basis states $\{\ket{l}\lsub{i}\ket{l'}\lsub{s}\}$}

The probability for the basis states $\{\ket{l}\lsub{i}\ket{l'}\lsub{s}\}$ that appear in the
coherent superposition of the states $\ket{\Psi}_1$ and $\ket{\Psi}_2$ can be measured directly.
This is done with selective coincidence measurements recorded with the detectors placed just behind
the double slit and it can be seen as a test of our theory, since one can infer from these probabilities
the amplitudes of the coefficients of the states really generated by the the arms of the interferometer. In these
measurements, the detector $D_{s}$ is kept fixed behind one slit while the other detector scans,
in the $x$ direction, the entire region of its double slit. Two measurements of this kind with detector
$D_{s}$ fixed behind the slit ``+'' ($l=+$) and then at the slit ``-" ($l=-$), allow the determination
of the probabilities for all the four basis states for the state of one of the interferometer's
arms \footnote{This type of measurement can also be done in the plane of image formation of
these apertures \cite{GLima2}.}. It should be clear that for measuring the amplitudes of the
coefficients of $\ket{\Psi}_1$, the arm 2 should be blocked and
vice-versa. To determine the coefficients of the other state, this procedure must be repeated.

\begin{figure}
\begin{center}
\includegraphics[width= 0.5\textwidth]{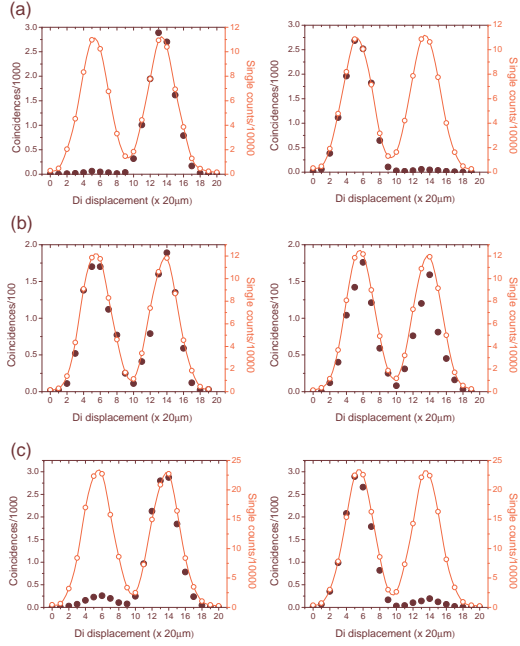}
\end{center}
\caption{$D_{i}$ single counts ($\circ$) and coincidence counts
($\bullet$) measured in 20~s, simultaneously, with $D_{s}$ fixed
behind one of the slits of its aperture. The data on the left side was recorded
with $D_s$ fixed behind the slit ``+" and on the right side with it fixed behind slit
``-". In (a) the measurements were done for determining part of the state of arm 1.
This was done with arm 2 blocked. In (b) they were done for determining part of the state of arm 2.
Arm 1 was blocked in this measurement. In (c) both arms are unblocked.}
\label{fig3}
\end{figure}

According to the quantum state in Eq.~(\ref{qubittomo}), one is expected to have peaks of
coincidences only when detector $D_{i}$ passes by the slit for which $l'=-l$. However,
for the state of arm 2 (Eq.~(\ref{prodtomo})), coincidence peaks should happen with
approximately the same number of coincidences between them, even when $l'\neq-l$. The
experimental data recorded is shown in Fig \ref{fig3}. One can clearly see that our
results are in agreement with our theoretical predictions.

The general expression for the states that are more ``likely" to have produced these
results are, for arm 1

\begin{eqnarray}        \label{phimes}
\ket{\Phi}_1 &= &0.077 \ket{++} + 0.704 e^{i\phi}
\ket{+-} \nonumber \\
& &+ 0.699 e^{i\phi} \ket{-+} + 0.099 \ket{--},
\end{eqnarray}
whose fidelity with the state $\ket{\Psi}_1$ (Eq~(\ref{qubittomo})) is $F = 0.98 \pm 0.05$. And for arm 2

\begin{eqnarray}        \label{phiparc}
\ket{\Phi}_2 &= &0.514 \ket{++} + 0.502 e^{i\phi}\ket{+-} \nonumber \\
& &+ 0.501 e^{i\phi}\ket{-+} + 0.483\ket{--},
\end{eqnarray} whose fidelity with the state $\ket{\Psi}_2$ is $F = 1.00 \pm 0.08$.

\subsection{The Measurement of $A$ and $B$}

The probabilities for generating the states $\ket{\Psi}_1$ ($A$) and $\ket{\Psi}_2$ ($B$)
while both arms of the interferometer are unblocked can also be measured.
And this measurement can also be used as a test for the theory given above.

We measured the values of $A$ and $B$ by blocking one of the arms
of the interferometer and detecting the transmitted coincident
photons through the signal and idler double-slits (See Fig \ref{fig3}). A (B) is the
ratio between the total coincidence when arm 2 (arm 1) is blocked
and the total coincidence when both arms are unblocked. From
this measurement we obtained $A = 0.85 \pm 0.03$ and $B = 0.15
\pm 0.03$. The reason for having the probability
of generating the state of arm 1 much
higher than the probability for generating the state of arm 2 is quite
simple. The laser beam that crosses arm 1 of the interferometer is
focused at the slits' plane and the spatial
correlation of the generated photon pairs is such that it is more favorable to their
transmission through the double slits than it is when the photon pairs are generated by the pump beam
that crosses arm 2. These values of $A$ and $B$ can be properly
manipulated by inserting attenuators at the interferometer.

\begin{figure}[t]
\begin{center}
\includegraphics[width=0.35\textwidth]{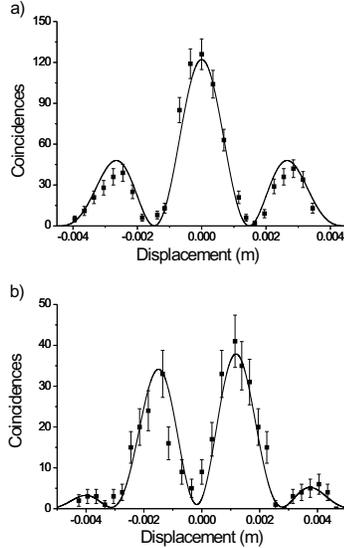}
\end{center}
\caption{Fourth order interference pattern as a function of
$D_{s}$ position. In (a), the detector idler was fixed at the
transverse position $x = 0$~mm. In (b), it was fixed at the
transverse position $x = 1376$~$\mu$m. The solid curves were obtained
theoretically.} \label{fig:Conditional}
\end{figure}

This result can now be used for a test of our theory. To do this we must consider the
detection of the fourth order interference patterns \cite{MandelBook} at a transverse
$z$-plane far behind the plane of the double slits. Since the state $\ket{\Psi}_1$ is
a maximally entangled state, we would expect to observe strong conditional interference
patterns \cite{Fonseca,Greenberger} for a mixture given by Eq~(\ref{OPDENSTOMO}) and
with a high value of $A$, when both interferometer arms are unblocked. This would not
be the case for high values of $B$, since the degree of entanglement of $\ket{\Psi}_2$
is at least three times smaller than the degree of $\ket{\Psi}_1$. The existence of a
relation between the conditionality of the fourth-order interference patterns and the degree of
entanglement of the two spatial qubit states was proven in Ref \cite{ConcEsp}. The interference
patterns were recorded at a transversal plane that was $600$~mm behind the double slit's
plane. They were recorded as a function of the $D_{s}$ $x$-position and they are shown in
Fig \ref{fig:Conditional}. In Fig \ref{fig:Conditional}(a), the idler detector was fixed at
$x = 0$~mm. In Fig \ref{fig:Conditional}(b), it was fixed at the transverse position
$x = 1376$~$\mu$m. The solid curves were obtained theoretically by using
Eq~(\ref{OPDENSTOMO}) and the measured values of $A$ and $B$. The conditionality can
be clearly observed in the interference patterns. One can also clearly see the good fit between the
theoretical curve and our results.

As we discussed in the beginning of this section, these experimental observations corroborate
the use of states $\ket{\Psi}_1$ and $\ket{\Psi}_2$ as good approximations for the states
generated through arm 1 and arm 2 of the interferometer used.

\section{Conclusion}

In conclusion, we have shown that it is possible to generate a broader family of composite
systems of spatial qudits by exploring the transverse correlations of the down-converted
photons and the effects of optical interferometry. The process was discussed in detail and
experimental evidences were shown that corroborate with the theory here proposed.

\section*{Acknowledgments}

The authors acknowledge the support of the Brazilian
agencies CAPES, CNPq, FAPEMIG and Mil\^enio Informa\c{c}\~ao Qu\^antica. C. Saavedra and A. Delgado were supported by
Grants Nos. FONDECYT 1061046 and Milenio ICM P06-67F. F. Torres-Ruiz was supported by MECESUP
UCO0209. This work is part of the international cooperation agreement CNPq-CONICYT
491097/2005-0.

\end{document}